\documentclass[aip,apl,amsmath,amssymb,reprint,]{revtex4-1}
\usepackage{graphicx}
\usepackage{epstopdf}
\usepackage{dcolumn}
\usepackage{bm}
\usepackage{CJK}
\usepackage{amsmath}
\usepackage{color}
\usepackage{soul}
\begin{document}

\title{Electro-Optic Lithium Niobate Metasurfaces}

\author{Bofeng Gao}
\affiliation{The Key Laboratory of Weak-Light Nonlinear Photonics, Ministry of Education, School of Physics and TEDA Applied Physics Institute, Nankai University, Tianjin 300071, P.R. China}

\author{Mengxin Ren}
\email{ren$_$mengxin@nankai.edu.cn}
\affiliation{The Key Laboratory of Weak-Light Nonlinear Photonics, Ministry of Education, School of Physics and TEDA Applied Physics Institute, Nankai University, Tianjin 300071, P.R. China}
\affiliation{Collaborative Innovation Center of Extreme Optics, Shanxi University, Taiyuan, Shanxi 030006, P.R. China}

\author{Wei Wu}
\affiliation{The Key Laboratory of Weak-Light Nonlinear Photonics, Ministry of Education, School of Physics and TEDA Applied Physics Institute, Nankai University, Tianjin 300071, P.R. China}

\author{Wei Cai}
\affiliation{The Key Laboratory of Weak-Light Nonlinear Photonics, Ministry of Education, School of Physics and TEDA Applied Physics Institute, Nankai University, Tianjin 300071, P.R. China}

\author{Jingjun Xu}
\email{jjxu@nankai.edu.cn}
\affiliation{The Key Laboratory of Weak-Light Nonlinear Photonics, Ministry of Education, School of Physics and TEDA Applied Physics Institute, Nankai University, Tianjin 300071, P.R. China}

\begin{abstract}
Many applications of metasurfaces require an ability to dynamically change their properties in time domain. Electrical tuning techniques are of particular interest, since they pave a way to on-chip integration of metasurfaces with optoelectronic devices. In this work, we propose and experimentally demonstrate an electro-optic lithium niobate (EO-LN) metasurface that shows dynamic modulations to phase retardation of transmitted light. Quasi-bound states in the continuum (QBIC) are observed from our metasurface. And by applying external electric voltages, the refractive index of the LN is changed by Pockels EO nonlinearity, leading to efficient phase modulations to the transmitted light around the QBIC wavelength. Our EO-LN metasurface opens up new routes for potential applications in the field of displaying, pulse shaping, and spatial light modulating.
\end{abstract}

\maketitle
Formed by artificial subwavelength building blocks known as meta-atoms, metasurfaces have demonstrated abilities to control optical waves with unprecedented flexibility and opened up recently our imagination for realizing a new generation of flat optical components outperforming their conventional bulky counterparts.\cite{zheludev2012metamaterials} Despite their impressive advances, current metasurfaces are mostly static in nature whose optical properties are set in stone after their fabrication process. Realizing modulation of the metasurfaces' properties in time domain can provide new opportunities to manipulate light and facilitate a transition to dynamic optical devices.\cite{li2017nonlinear, shaltout2019spatiotemporal, he2019tunable, ren2020tailorable} For this purpose, different dynamic tuning mechanisms such as optical pumping,\cite{wang2016optically, taghinejad2018ultrafast, ren2017reconfigurable, nicholls2017ultrafast,guan2019pseudospin} thermal heating,\cite{ou2011reconfigurable} chemical reaction,\cite{di2016nanoscale,duan2017dynamic} and electrical stimulation\cite{samson2010metamaterial,karvounis2020electro} have been implemented. Among all these tuning mechanisms, electrical tuning techniques are of particular interest, because they hold a promise to integrating the metasurfaces with other on-chip optoelectronic devices. The most common electrical methods reported so far are based on triggering free carrier modulations,\cite{salary2020tunable,chen2006active, dabidian2015electrical} molecular reorientations,\cite{decker2013electro} and phase transitions\cite{driscoll2009memory,jia2018dynamically} in active materials integrated in the meta-atoms. However, above approaches rely on relative slow physical processes and the switching time is normally limited below nanosecond.\cite{lee2014ultrafast}

Lithium niobate (LN) is one of the most appealing materials to overcome this challenge. The LN shows outstanding Pockels electro-optic (EO) effect, and their refractive index could be changed by an electrical voltage within a femtosecond timescale.\cite{gaborit2013nonperturbative} Thus the LN enables optical modulators with much higher switching rates.\cite{el2008optical} Recently, thin-film LN on insulator (LNOI)\cite{levy1998fabrication,rabiei2004optical} emerges as a promising platform for ultracompact  photonic devices.\cite{poberaj2012lithium,fang2017monolithic,qi2020integrated,kong2020recent} Thanks to the large refractive index contrast between the LN film and the substrate (such as silica), optical modes are tightly confined within the nanometers-thin LN layer leading to improved EO modulation efficiency. And a variety of on-chip EO modulator units with tens to hundreds of GHz modulation speeds have been demonstrated using different LNOI microstructures, for example Mach-Zehnder interferometric waveguide,\cite{wang2018integrated,he2019high} photonic crystals,\cite{ding2019integration,li2020lithium} micro-rings\cite{guarino2007electro,zhang2019broadband} or micro-disks.\cite{bo2015lithium} In recent years, there have been significant advances in fabricating LN metasurfaces\cite{gao2019lithium,fang2020second}, which led to the demonstration of intriguing tunable second harmonic properties.\cite{ma2020resonantly,fedotova2020second} However, the EO modulation by the LN metasurface, to the best of our knowledge, has not been experimentally explored.

The large EO modulation strength essentially implies sensitively tunable metasurfaces' properties (such as phase retardation) by the EO induced refractive index changes. For this purpose, an efficient way is to utilize high-quality factor (high-Q) resonant modes with a narrow spectra linewidth which significantly elongate the effective optical path and photons lifetime in the meta-atoms, yielding enhanced local fields that experience the EO-refractive index changes. An attractive approach for the extremely high Q-factors is provided by the physics of bound states in the continuum (BICs). Such concept was first proposed in quantum systems where the electron wave function exhibits localization within the continuous spectrum of propagating waves.\cite{von1929remarkable,friedrich1985interfering} Recently, BICs have also attracted considerable attentions in photonics.\cite{marinica2008bound,hsu2016bound} Mathematically, the BIC states show infinite Q-factors where the optical energy is trapped without leakage.\cite{koshelev2018asymmetric,xiang2020tailoring} The ideal BICs have a vanishing spectral linewidth and are not observable in the electromagnetic spectra. However, in practice, introducing a structural asymmetry or oblique excitation could break the ideal BIC conditions.\cite{liu2019high,huang2020highly} Consequently the perfect BIC modes will convert to quasi-BIC (QBIC) states that manifest themselves as extremely narrow spectral resonances with large Q-factors.\cite{azzam2018formation,han2019all} Such QBICs have been observed in extended photonic systems and hold a great promise for various applications including vortex beam generation,\cite{wang2020generating} nonlinear enhancement,\cite{carletti2018giant,krasikov2018nonlinear,anthur2020continuous} low threshold lasing\cite{kodigala2017lasing,huang2020ultrafast} and sensitive sensing.\cite{romano2018label,leitis2019angle}

\begin{figure}[tphb]
\includegraphics[width=75mm]{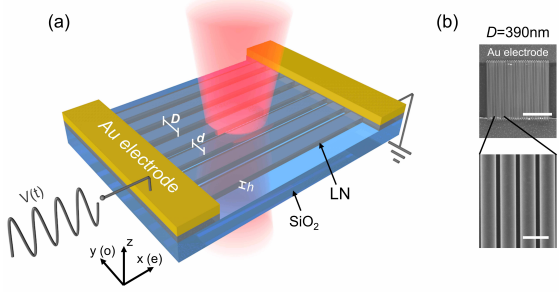}\caption{\label{fig1}
\textbf{Design of EO-LN metasurface.} (a) Schematic illustration of the LN metasurface. The properties of the meta-atoms are actively modulated by applying an external electric voltage V(t). The geometrical parameters of the metasurface are lattice constant $D$, ridge width $d$, and thickness $h$. (b) Scanning electron microscope images of the fabricated sample. Scale bars are 5~$\mu$m (up panel) and 500~nm (down panel), respectively.}
\end{figure}

In this paper, we numerically and experimentally demonstrate a LN metasurface offering EO phase modulation to transmitted light in the visible frequency regime. To yield an obvious phase modulation, we utilize a nanograting array under oblique incidence in which the EO effect induced by applied bias voltage is significantly enhanced by leveraging the QBIC states, resulting a 1.46~times larger EO modulation strength compared with the unstructured LN film. To the best of our knowledge, this is the first experimental demonstration of the EO modulation by the LN metasurfaces. Our results would act as a novel dynamic EO platform for wavefront engineering, pulse shaping, and polarization control, etc.

A designed schematic of the EO-LN metasurface is shown in Fig.~1. The metasurface is composed of an array of LN nanogratings residing on a fused quartz substrate. The LN ridge width is presented by $d$ and grating period by $D$. Height $h$ is 200~nm which is determined by thickness of the LN film (LNOI by NANOLN corporation) used for the metasurface fabrication. In our design, the orientation of the nanogratings is parallel to the $e$-axis of the LN crystal. And the external electric field is applied also along the $e$-axis to take the advantage of the largest EO-coefficient element $\gamma_{33}$. We fabricated the metasurfaces by focused ion beam technique (FIB, Ga$^+$, 30kV, 24pA) following previous procedure.\cite{gao2019lithium} And footprint of the metasurface array is 10$\times$10$\mu$m$^2$. The Au electrodes with 10~$\mu$m gap were fabricated via standard UV-lithography procedure. Figure~1 (b) gives the scanning electron microscope (SEM) images of total footprint and zoomed-in view of the fabricated LN-EO metasurface with $D$=390~nm and $d$=290~nm.

\begin{figure}[phb]
\includegraphics[width=85mm]{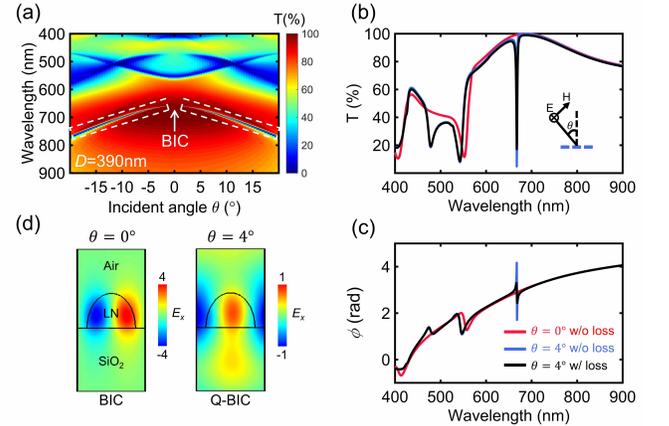}\caption{\label{fig2}
\textbf{BIC states in LN metasurface.} (a) Incident-angle resolved transmission spectra of the LN metasurface. (b) Comparison of transmission $T$ between 0$^\circ$ (red line) and 4$^\circ$ (blue and black lines) incidence shows a collapse of BIC to sharp Fano shaped QBIC resonance. The incident wave is polarized along $x$-axis and located within $yoz$-plane. The LN is assumed to be lossless for red and blue lines, while $n_i$ is set as 0.002 to consider the loss caused by FIB fabrication (black line). (c) Spectra for transmitted phase $\phi$. (d) Electric field distributions of eigenmodes for 0$^\circ$ and 4$^\circ$ incidence. The electric fields are tightly confinement within LN layer at 0$^\circ$ and shows a clear leakage to substrate for 4$^\circ$ incidence.}
\end{figure}

Figure~2(a) shows a full map of the transmission spectrum (T) of the metasurfaces with $D$ = 390~nm as a function of the incident angle ($\theta$) under $x$-polarized incident light. The spectra were calculated using a finite element method (COMSOL Multiphysics). And ellipsometric measured refractive indices of the birefringent LN and the fused quartz were used in the simulations. Such grating structures are expected to support the symmetry-protected BIC modes at normal incidence.\cite{hsu2016bound} And as shown in Fig.~2(a), resonant modes indicated by white dashed rectangles clearly emerge for oblique incidence (nonzero $\theta$). In order to clarify the behavior of the BIC modes, we plot the transmission T and phase retardation ($\phi$) spectra in Fig.~2(b) and (c), respectively. It can be observed that ultra-narrow asymmetric Fano-shaped transmission dips and abrupt phase slips occur around the wavelength of 667~nm for incident angle of $\theta$=4$^\circ$ (blue lines). These resonances vanish from the spectra at $\theta$=0$^\circ$ (red curves). Such characteristic is clear a manifestation of occurrence of the BIC resonant modes. Left panel of Fig.~2(d) demonstrates the eigenmode distribution of the $x$-component of electric field ($E_x$) at $\theta$=0$^\circ$ in the $yoz$ cross section of the meta-atom. The mode exhibits an antisymmetric profile along the horizontal direction with a node formed at the center, corresponding to an odd mode parity symmetry. The electromagnetic fields are tightly confined in the LN layer and decoupled from the free-space propagating waves. However, for $\theta$=4$^\circ$ the electric energy clearly leaks out into SiO$_2$ substrate, and the magnitude of the electric fields in the LN  become 4 times weaker than the ideal BIC mode. Such phenomena further confirm the presence of the true BIC for normal incidence which collapses into the QBIC modes for oblique excitation. Despite the LN is ideally lossless within the studied spectral range, however the Ga$^+$ contamination and the lattice damage during the FIB milling will inevitably deteriorate the optical performance of the metasurface.\cite{geiss2014photonic} Such influence was taken into account in the simulation by putting the imaginary part of the LN refractive index $n_i$ as 0.002. And the calculated results are shown by black curves. It can be clearly seen that the ultra-sharp dip in $T$ and abrupt phase slip in $\phi$ are preserved, however the resonance strength in both $T$ and $\phi$ is reduced in the presence of optical loss.

Such high Q-factor QBIC resonance leads to a increased lifetime of photons and strong localization of the fields within the meta-atoms, which would significantly enhance the light-matter interaction at nanoscale and boost the spectral tunability resulting from the EO induced refractive index change in the LN. And the extremely sharp QBIC phase resonance can yield a substantial phase modulation in transmission through small EO spectral shifts of the modes. Figure~3 demonstrates the simulated phase spectra of the transmitted light through the metasurface for different variations in the real part of refractive index of the LN while maintaining $n_i$=0.002. It can be clearly observed that the phase spectrum shifts by 0.6~nm to shorter wavelengths for the reduced refractive index, while redshifts by the increased refractive index. The choice of operating wavelength at the QBIC resonance wavelength of 667~nm is denoted by the vertical dashed line in Fig.~3(a). The phase modulation $\Delta\phi$ at this wavelength are calculated as a function of $\Delta n$ and the results are plotted in Fig.~3(b). It is shown that a modulation span $\Delta\phi$ of $\pm$0.42 rad in the transmitted light phase is obtained through tuning QBIC resonance via a $\Delta n$ modulation from -0.0025 to 0.0025.

\begin{figure}[phb]
\includegraphics[width=65mm]{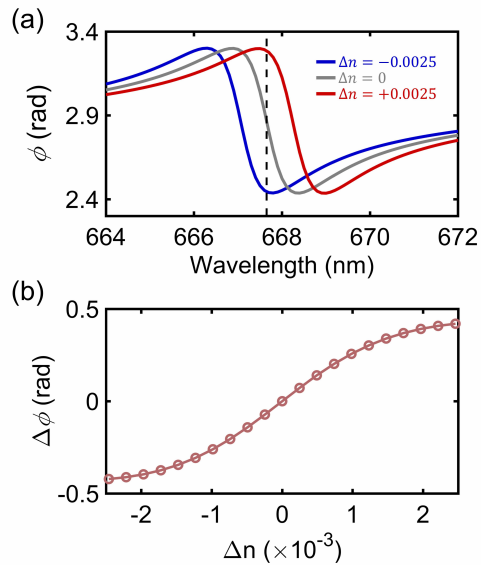}\caption{\label{fig3}
\textbf{Simulated EO-phase modulation by metasurface.} (a) Transmission phase spectra of the LN metasurface with $D$=390~nm around QBIC wavelength (indicated by vertical dashed line) for 4$^\circ$ oblique incidence corresponding to different refractive index modulations. (b) Relation between transmitted phase modulation and refractive index modulation for the LN metasurface at the QBIC wavelength.}
\end{figure}

To experimentally evaluate the transmission spectrum of the fabricated LN metasurface, we built a micro-spectrometer. Output of a supercontinuum laser (NKT EXR-15) was focused onto the LN metasurfaces by a 10$\times$ objective. The transmitted light was analyzed using a spectrometer (Horiba MicroHR). The measured transmission spectrum under $x$-polarized incidence is given in Fig.~4(b). A QBIC resonance dip clearly appears around 633~nm, as indicated by a vertical black arrow. It is shown that the experimentally measured QBIC resonance is much shallower and broader leading to a smaller Q-factor compared with the simulation results shown in Fig.~2. Such discrepancy could be explained by the fabrication imperfections. Furthermore, both normal and oblique incidence components were included in the experiment, thus the QBIC dip may be averaged out by the normal incident component.

\begin{figure*}[tphb]
\includegraphics[width=120mm]{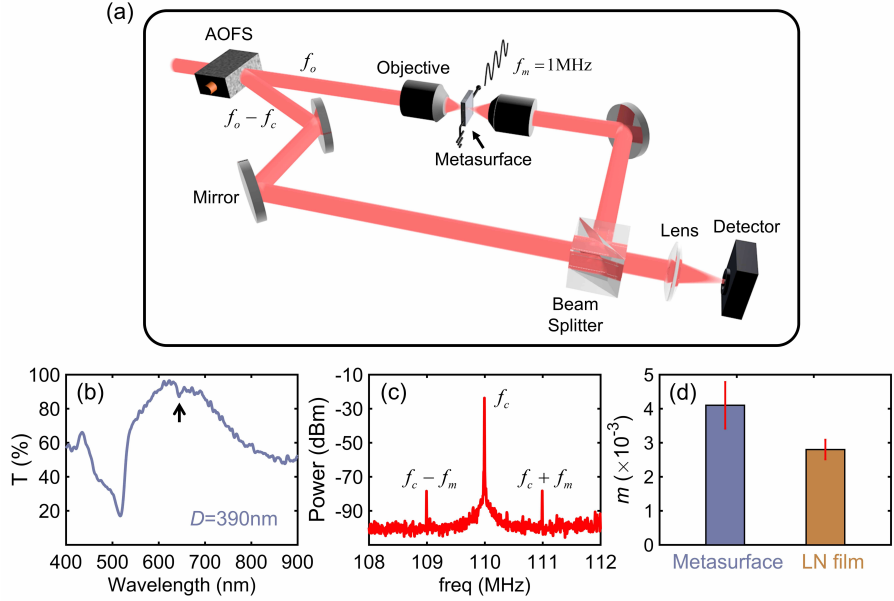}\caption{\label{fig4}
\textbf{Experimental EO phase modulation by metasurface.} (a) A schematic diagram of a laser heterodyne detection system. An acoustic-optic frequency shifter (AOFS) is used to divide an input 633~nm laser (with optical frequency of $f_o$) into two parts. The 0th order light is used as probe light to excite the LN metasurface. And the frequency of the 1st order light is downshifted by $f_c$=110~MHz and used as reference light. The metasurface is modulated by a sinusoidal electric voltage at $f_m$=1.0~MHz. (b) Experimentally measured transmission spectrum of the metasurface with $D$=390~nm excited by focused light. A clear QBIC resonance dip is observed around 633~nm as indicated by a vertical black arrow. (c) Power spectrum of optical beats recorded by a RF spectrum analyzer. Three distinct peaks are observed at $f_c-f_m$=109~MHz, $f_c$=110~MHz and $f_c+f_m$=111~MHz. (d) Phase modulation magnitude $m$ measured from the LN metasurface and the unstructured LN film. Heights of histograms are average value of multiple measurements, and error bars are their standard deviations.}
\end{figure*}

The EO phase modulation by the metasurfaces was characterized by a home-built laser heterodyne detection system [shown in Fig~4(a)]. A $x$-polarized 633~nm continuous wave laser (CNILaser, MGL-III-532) was launched into an acoustic-optic frequency shifter (AOFS) and was divided into two parts, i.e. the 0th order transmitted and the 1st order diffracted light beams. The 0th order light without frequency shift was used as probe light to focus onto the LN metasurface by the 10$\times$ objective. And the 1st order light with a frequency downshift of $f_c$=110~MHz was used as reference light. An arbitrary waveform generator (Agilent 33250A) was used to generate a sinusoidal driving voltage signal at $f_m$=1.0~MHz, which was further amplified to be 300~V$_{pp}$ (peak-to-peak magnitude, -150V to +150V output voltage) using a high-voltage amplifier (Falco Corp.), and then was fed into the electrodes. As a consequence, the phase of the probe light was changed by the EO response of the LN metasurface. After interfering the probe light with the reference beam, optical beats were generated which were further recorded by a photodetector and a RF spectrum analyzer. In our measurements, the visibility of the optical beats has been optimized by equilibrate the powers and optical paths of the two beams.

Assuming the probe and reference beams at the photodetector are $E_p=e^{i[2\pi f_ot+m\sin(2\pi f_mt)]}$ and $E_r=e^{i2\pi (f_o-f_c)t}$, respectively where $f_o$ is the optical frequency of the 633~nm laser, and $m$ is the EO phase modulation magnitude. Thus the optical beats can be described as $I=|E_p+E_r|^2=2+2\cos[2\pi f_c t+m\sin(2\pi f_m t)]$. When the $m\ll1$, such beat signal could be expressed by a set of standard Bessel function expansions $I=2+2\{J_0(m)\cos(2\pi f_ct)+J_1(m)[\cos(2\pi(f_c+f_m)t)-\cos(2\pi(f_c-f_m)t)]\}$. And the corresponding Fourier frequency spectrum can be expressed as $\mathcal{F}(f)=2\delta(f)+J_0(m)\delta(f-f_c)+J_1(m)[\delta(f-(f_c+f_m))-\delta(f-(f_c-f_m))]$, in which $\delta(f)$ is the Kronecker delta function.\cite{zhang2015optical} This equation indicates that the phase modulation signal results in three discrete frequency components $f_c-f_m$, $f_c$ and $f_c+f_m$, respectively. And as shown in Fig.4(c), indeed three distinct peaks are observed at 109~MHz, 110~MHz and 111~MHz in the experimental power spectrum. The magnitude of the frequency component at $f_c$ and ${f_c} \pm {f_m}$ are proportional to $J_0^2\left( m \right)$ and $J_1^2\left( m \right)$ respectively. Then the phase modulation magnitude $m$ can be mathematically demodulated by the experimental ratio of ${{J_0^2\left( m \right)} \mathord{\left/{\vphantom {{J_0^2\left( m \right)} {J_1^2\left( m \right)}}} \right.
 \kern-\nulldelimiterspace} {J_1^2\left( m \right)}}$. And the deduced $m$ for the metasurface and the unstructured LN film are shown in Fig.~4(d). The average $m$ value of 0.0041~rad is achieved from the metasurface, which is larger than the 0.0028~rad by the unstructured LN film. This explicitly show that the adoption of the metasurface is a valid way for stronger EO-modulation.

In conclusion, We have numerically and experimentally demonstrated an EO tunable LN metasurfaces that provide a dynamic control over the phase retardation of transmitted light in the visible spectral regime. The oblique incidence enables collapse of symmetry-protected BICs into Fano resonant QBIC modes with ultrahigh Q-factors, which significantly increases the lifetime of photons and field confinement within the resonators leading to the improved modulation sensitivity. The proposed EO-LN metasurface is of great interest for developing multifunctional and tunable optical components such as ultracompact spatial light modulators, optical switches, which would find users in various applications including displaying, optical wavefront shaping, and so on.

\begin{acknowledgments}
This work was supported by National Key R\&D Program of China (2017YFA0305100, 2017YFA0303800, 2019YFA0705000); National Natural Science Foundation of China (92050114, 91750204, 61775106, 11904182, 12074200, 11774185); Guangdong Major Project of Basic and Applied Basic Research (2020B0301030009); 111 Project (B07013); PCSIRT (IRT0149); Open Research Program of Key Laboratory of 3D Micro/Nano Fabrication and Characterization of Zhejiang Province; Fundamental Research Funds for the Central Universities (010-63201003, 010-63201008, and 010-63201009); Tianjin youth talent support program. We thank Nanofabrication Platform of Nankai University for fabricating samples.
\end{acknowledgments}


\end{document}